\documentclass[aps,pre,floatfix,twocolumn,showpacs,showkeys,amsmath,amssymb,superscriptaddress,longbibliography,10pt]{revtex4-1}

\usepackage{amsthm}
\usepackage{bm}
\usepackage{epsfig}
\usepackage{verbatim}
\usepackage{graphicx}
\usepackage[normalem]{ulem}

\newcommand{\beq}{\begin{equation}}
\newcommand{\eeq}{\end{equation}}
\newcommand{\beqa}{\begin{eqnarray}}
\newcommand{\eeqa}{\end{eqnarray}}

\usepackage[table]{xcolor}

\begin{document}

\title{Barriers, trapping times and overlaps between local minima in the 
  dynamics of the disordered Ising $p$-spin Model}

\author{Daniel A. Stariolo}
%\email{stariolo@if.uff.br}
\affiliation{Universidade Federal Fluminense, Departamento de F\'isica and 
  National  Institute of Science and Technology for Complex Systems, Av. Gal. Milton Tavares de Souza s/n, 
  Campus da Praia Vermelha, 24210-346 Niter\'oi, RJ, Brazil}
\author{ Leticia F. Cugliandolo}
\affiliation{Sorbonne Universit\'e, Laboratoire de Physique Th\'eorique et Hautes Energies, 
  UMR 7589 CNRS, Tour 13, 5\`eme Etage, 4 Place Jussieu, 75252 Paris 05, France and \\
  Institut Universitaire de France, 1 rue Descartes, 75231 Paris Cedex 05, France}

\date{\today}
						
\begin{abstract}
  We study the low temperature 
  out of equilibrium Monte Carlo dynamics of the disordered Ising $p$-spin Model with $p=3$ and 
  a small number of spin variables. We focus on sequences of configurations that are 
  stable against single spin flips obtained by instantaneous gradient descent from persistent ones. 
  We analyze the statistics of energy gaps, energy barriers and trapping times on sub-sequences 
  such that the overlap between consecutive configurations does not overcome a 
  threshold. We compare our results to the predictions of various Trap Models finding the best agreement 
  with the Step Model when the $p$-spin configurations are constrained to be uncorrelated.
\end{abstract}

%{Statistical mechanics of model systems}
%{Metastable phases}
%{Applications of Monte Carlo methods}
%\pacs{05.50.+q,64.60.Cn,75.40.-s}

\maketitle

\section{Introduction}
\label{sec:introduction}

The Ising Model with disordered interactions between groups of $p$-spins, called the $p$-spin Model,
is one of the best understood systems with quenched randomness~\cite{Derrida1981,Gardner1985,Stariolo1990,Montanari2003,Rizzo2013a}.
Two paradigmatic models of spin glasses and glasses, the Sherrington-Kirckpatrik (SK) Model \cite{SK1975} and
the Random Energy Model (REM)~\cite{Derrida1980,Gross1984}, are limiting cases 
%of the $p$-spin Model when 
in which $p=2$, and $p \to \infty$ taken after $N\to\infty$, respectively. 

When $p \geq 3$ many static and dynamic
properties are in striking analogy with the ones of glass forming liquids~\cite{KiTh1987b,KiTh1987a}.
By rendering the spins continuous and imposing a global spherical constraint, the equilibrium properties of the
resulting ``spherical $p$-spin Model'' can be
solved exactly in the thermodynamic limit~\cite{Crisanti1992,Castellani2005}. Furthermore, the long time dynamic equations that couple
correlations and linear response functions have been shown to be equivalent to the 
Mode Coupling Theory (MCT)
equations for supercooled liquids and glasses also in the thermodynamic limit~\cite{CuKu1993,Bouchaud1996,Gotze2008}.
Both MCT equations and the spherical $p$-spin Model exhibit a dynamical
singularity at a temperature $T_d$, where relaxation times diverge. This singularity
is an artefact of the mean-field character of the MCT and $p$-spin Models. In finite dimensional systems
the dynamical transition should be in fact a crossover at $T_g$~\cite{Berthier2019} while the putative glass transition
should happen at a lower temperature $T_K$. Moreover, the original MCT approach has to be replaced
by a refined one that allows one to deal with the dynamics of finite systems, or infinite systems with short-range interactions~\cite{Rizzo16,Rizzo20}.

Further valuable information comes from
studies of the topological properties of the potential energy landscape (PEL) of model systems. 
Analytic and numerical results for the spherical and Ising $p$-spin Models
~\cite{Rieger1992,Crisanti1995,Cavagna1998,Crisanti2005a,Mehta2013,Ros2018a} and numerical results
for model glass formers~\cite{Angelani2000,Grigera2002,Angelani2003,Cavagna2009}
show that the number of stationary states (saddles)  of the PEL grows exponentially with system size $N$.
More importantly, when approaching $T_d$ from above 
%(or, equivalently, a threshold energy density $e_{th}$~\cite{CuKu1993}),
the number of unstable directions of typical saddles decreases strongly, and evidence has been presented
  that the MCT transition in finite dimensional glass formers
  corresponds to a localization transition of the unstable modes~\cite{Coslovich2019}.
At least in mean field models, below $T_d$, 
minima are exponentially more numerous than higher order saddles, 
and activation over barriers should be the dominant mechanism for relaxation.
Nevertheless, because of the mean-field character of
the $p$-spin Model, barrier heights diverge with $N$ and activation is suppressed, giving rise to the sharp
dynamical transition.

Describing the dynamical processes  of finite range interacting or finite size mean-field models 
below the crossover temperature, $T_g$ or  $T_d(N)$, is a great challenge. In the
nineties, the analysis was initiated with  studies of completely connected (mean-field) 
models of small size, so as to have some control over the barrier heights~\cite{Crisanti2000c,Crisanti2000a}.
Interesting results were obtained connecting the equilibrium behaviour of these models with
the ``metabasin'' concept introduced in studies of glass forming liquids~\cite{Stillinger1995,Buchner2000,Doliwa2003}. 

Another route to study activated
dynamics in disordered systems came from the so-called 
``Trap Models''~\cite{Dyre1987,Bouchaud1992}. These are
toy models, which stochastic dynamics can be exactly solved, defined by a set of states with
uncorrelated random energy levels.
In order to go from a trap to another the system has to jump over a  
barrier, spending a 
trapping time to do it that is given by an Arrhenius law. Given the distribution of trap energies and 
trapping times for each trap, it is possible to predict the form of the distribution of mean trapping times 
and the exact form of two-times correlation functions, the ``Arcsin law''~\cite{BouchaudDean1995,Monthus1996}. 
The Trap Model has become a paradigm for activated slow dynamics in disordered systems, due mainly to its
simplicity and the qualitative resemblance with the phenomenology of more realistic models, like the
$p$-spin and other model glass formers~\cite{Denny2003,Doliwa2003,Cammarota2018,Stariolo2019}. 
Nevertheless, a question that remains still open is to what extent the quantitative predictions of the
Trap Model are universal. In recent years much progress has been made in this direction. 
In the context of
the $p$-spin family of models, a series of rigorous results showed the validity of the Arcsin law for the
correlations in the aging regime of the REM ($p \to \infty$ after $N\to\infty$)
under certain conditions on the time scales of 
observation~\cite{Arous2002,BenArous2003,Bovier2003,Gayrard2016,Cerny2017}. These results rely strongly on the
independence of the random energy levels of the REM, which assures the renewal character of the dynamics
at each time step, similar to the updates in the Trap Models.
Extensions to  finite $p$, where energies are correlated, though still within 
simplified microscopic dynamics, have also been considered~\cite{Arous2008a,Bovier2013} confirming the
universal character of the Trap Model correlations. From the numerical
side, simulations of the single-spin-flip dynamics in the REM showed evidence for the Trap Model
scenario although they proved to be tricky to interpret~\cite{Baity-Jesi2018a,Baity-Jesi2018b}.
Further confirmation of the validity of the Arcsin law 
came from analytic and numerical work on extensions of the original Trap Model dynamical rules
in which the transition rates depend both on the initial and final 
states~\cite{Rinn2000,Rinn2001,Cammarota2018}. 
In particular, 
Glauber~\cite{Barrat1995,Bertin2003} and Metropolis~\cite{Cammarota2015} microscopic updates were used and lead 
to the so-called  ``Step Model''.
This model shares the distribution of random energies of the original Trap Model, 
but the Glauber or Metropolis dynamics
may lead to relaxation without the need for activation over energy barriers. In fact, at sufficiently
low temperatures, the kind of ``entropic'' activation promoted by the Metropolis rule is the only
relevant one for relaxation, leading to complex aging dynamics similar to the ones of the Trap Models.
Interestingly, subsequent work showed that at intermediate temperatures energetic activation is also
at work, leading to a competition between energetic and entropic mechanisms in the relaxation~\cite{Bertin2003}. 
Furthermore, detailed numerical studies showed that the Arcsin law also emerges in
the dynamics of the Step Model after a suitable coarse-graining of the landscape, leading to a redefinition
of the traps as sets of configurations, or basins, rather than single configurations~\cite{Cammarota2015}.
It is important to note that all the evidence obtained so far in favor of a ``Trap Model universality''
  for aging dynamics in disordered systems relies on a very strong assumption,
  namely, that that the dynamics is a {\em renewal process}. In such a process, the evolution in time does not depend on
  the past history. This is a strong assumption, the general validity of which is not guaranteed
  \cite{Sibani2013,Boettcher18}.

Here, we wish to address whether the quantitative
predictions of the Trap and Step Models can emerge in the dynamical behaviour of the standard 
$p$-spin Ising Model. We consider the latter with $p=3$ and single-spin-flip Metropolis 
updates. From its known properties,
we may expect the $p$-spin Model to show some characteristics of both the Trap and Step Models, namely
energetic and entropic relaxation mechanisms. A first challenge is related to finite size effects.
In Ref.~\cite{Stariolo2019} we showed that, in order to access the time scales relevant to activation,
it is necessary to consider rather small systems. A second challenge is the very definition of
``trap'' in the context of the $p$-spin Model. In \cite{Stariolo2019} we
proposed an ad-hoc definition of trap, based on the observation of persistent configurations in the
low temperature dynamics of the model. That definition led to some interesting observations, like the
emergence of an exponential regime for the trap energies in the low energy sector of the landscape
and power law distributions of trapping times. Nevertheless, at a quantitative level, we were
unable to find a neat connection with the Trap Model, nor with the Step Model predictions.
In the present work, we refine the definition of trap, relating them to the presence of configurations
stable against single spin flip, which seem to be natural candidates for persistent states,
resembling the ``inherent structures'' classification in model glass formers~\cite{Stillinger1983,Crisanti2000a,Fabricius2004}. 
Furthermore, in order to tackle the problem of the strongly correlated energy levels
in the $p$-spin Model, we follow selected sequences of locally stable configurations which obey constraints in
the mutual overlap. In this way, we were able to consider sequences of configurations with different
degree of correlation. We performed a thorough characterization of the energy landscape visited by
the system during the Metropolis dynamics, and then used this information to analyze the results for
the distributions of trap energies and trapping times. In order to conform more closely with the definitions
of the Trap Model, in this work we defined traps as energy barriers between consecutive locally stable
configurations satisfying the overlap constraints. We compared our results with the predictions
of the original Trap Model, with the Trap Model with generalized dynamics and with the Step Model.
We found qualitative similarities with all three models of trap dynamics but, interestingly, the
numerical results are in quantitative agreement only with the Step Model predictions 
in the limit of uncorrelated sequences of traps. 

The structure of the paper is the following. In Sec.~\ref{sec:model} we recall the definition of the 
Ising $p$-spin Model. We present the methodology in Sec.~\ref{sec:methodology}.
Section~\ref{sec:results} is devoted to the presentation of our numerical results
and Sec.~\ref{sec:comparison} to the comparison to the predictions of the 
Trap and Step Models. Finally, we close the paper with a discussion presented 
in Sec.~\ref{sec:discussion}.

\section{The $p$-spin Model} \label{sec:pspin}
\label{sec:model}

The Ising spin glass with multispin interactions is defined by the energy function:
\beq \label{eq:phamilton}
E  = -\frac{1}{p!}\sum_{i_1, i_2,\dots, i_p = 1}^{N} J_{i_1, i_2, \dots, i_p} S_{i_1}S_{i_2}\cdots S_{i_p}
\; ,
\eeq
where $\{S_i=\pm 1, i=1\ldots N\}$ are Ising spin variables and the coupling constants $J_{i_1, \dotsc, i_p}$
are quenched Gaussian random exchanges  with zero mean and standard deviation $\sigma=\sqrt{p!/2N^{p-1}}$.
The Hamiltonian \eqref{eq:phamilton} consists of $p$-spin interactions between all possible groupings of
different spins on $p$ sites. It is a fully connected model. The tensor of coupling constants
$J_{i_1, \dotsc, i_p}$ is symmetric under arbitrary permutations of the indices $\{i_1, i_2, \dots, i_p \}$.

The energies of single configurations are Gaussian random
variables with $P(E) \sim \exp{(-E^2/N)}$.
Furthermore, the probability that two configurations $S_1$ and $S_2$ have energies
$E_1$ and $E_2$ is given by
\beq \label{eq:enercorrels}
P(E_1,E_2) \sim \exp{\left[-\frac{(E_1+E_2)^2}{2N(1+q^p)}-\frac{(E_1-E_2)^2}{2N(1-q^p)}\right]}
\; ,
\eeq
at leading order in $N$~\cite{Derrida1981}.
Thus, pairs of configurations are correlated. As seen in Eq.~(\ref{eq:enercorrels})
the degree of correlation depends on
their overlap 
\begin{equation}
q(S^1,S^2)=\frac{1}{N}\sum_i S_i^1 S_i^2
\; , 
\end{equation}
with $|q|\leq 1$.
Having already taken the large $N$ limit, one can now take  $p \to \infty$
and find that different energy levels ($q<1$) become uncorrelated,
%\sout{\blue{In the large $N$ limit, after sending $p \to \infty$, different energy levels ($q<1$) become uncorrelated}}, 
$P(E_1,E_2) \sim P(E_1)P(E_2)$.
This limit corresponds to Derrida's Random Energy Model (REM)~\cite{Derrida1981} (see also a discussion in Ref.~\cite{Baity-Jesi2018b},
relevant to the finite $N$ situation).

\section{Methodology}
\label{sec:methodology}

We performed single-spin-flip Monte Carlo simulations 
of the completely connected Ising $p$-spin Model defined in
Eq.~(\ref{eq:phamilton}) with $p=3$~\footnote{We expect that the computationally more efficient
  finite connectivity model would yield essentially the same conclusions for the present study.
  Nevertheless, we decided to stick to the completely connected version to keep strict quantitative
  continuity with our previous related work \cite{Stariolo2019}}.
We worked with the Metropolis transition rates
from configuration $i$ to configuration $j$:
\begin{eqnarray}
r_{i,j} =
\left\{
\begin{array}{ll}
\tau_s^{-1} \, e^{-\beta \Delta E} & \qquad \Delta E >0 
\\
\tau_s^{-1} & \qquad \mbox{otherwise}
\end{array}
\right.
\label{eq:Metropolis}
\end{eqnarray} 
where $\Delta E = E_j - E_i$.
As usual, the time step unit is set to be the Monte Carlo step (MCs), $N$ flip attempts.
In all cases the system was prepared in a disordered initial state
and suddenly quenched to a low temperature. Because time scales for activation are
expected to grow exponentially with system size, we fixed $N=20$, which implies a configuration
space of $2^N\sim 10^6$ states. Typical simulations were run for a total time of up to $10^7$ MC steps. 
The temperature was fixed to $T=0.2$, much lower than both the dynamical temperature, 
$T_d= 0.682$, and the static critical temperature, $T_s=0.651$, of the infinite 
size system (although for $N=20$ the transitions are considerably
rounded~\cite{Billoire2005}). At this low temperature the system remains relaxing out
of equilibrium until the longest simulation times considered here, $t = 10^8$ MCs.
Statistics were recorded for a number of disorder samples between $3\times 10^4$ and $10^5$.

As discussed in~\cite{Stariolo2019}, during the evolution at low temperature the system eventually remains
trapped in single configurations for a certain number of MCs, until thermal fluctuations restore the evolution.
Those configurations were chosen as an essential part of the definition of
dynamical ``traps'' in our previous work~\cite{Stariolo2019}. Instead, in the present study we
followed sequences of configurations which are {\em stable against single spin flips}. These
configurations are analogs of the ``inherent structures'' defined as mechanically stable configurations
of the landscape in continuous models for the glass transition~\cite{Stillinger1983,Doliwa2003}.
Operationally, along
the dynamics we identify configurations which persist for at least 5 MCs \footnote{We only look at the
configurations at the end of each Monte Carlo step, i.e. we do not look at possible reversible single
spin flips within each MCs.}
and we define the following
quantities, which will be the base for our analysis:
\begin{itemize}
  \item {\em Locally Stable Configurations} (LSC): Once a persistent configuration
is identified, we perform a quench to zero temperature to reach the nearest single-spin-flip-stable
configuration, i.e. a ``locally stable configuration''.
\item {\em Barriers}: barrier heights are defined as the difference between the energy of a
  LSC and the maximum energy reached along the dynamical path before arriving at the next LSC.
  The configuration with the maximum energy between two successive LSC is called a ``transition
  state''~\footnote{Note that our definition of transition state does not
      coincide with the 
  one commonly used in studies of continuous energy landscapes, see. e.g. D. J. Wales, 
  {\it Energy landscapes: applications to clusters, biomolecules and glasses} (Cambridge Univ. Press, 2003).
  Here, transition states are not order one saddles connecting two minima.}.
\item {\em Gaps}: we define a ``gap'' as the energy difference between two consecutive LSC.
\item {\em Overlaps}: overlaps between two locally stable configurations, $\{S^1\},\{S^2\}$,
  are defined as usual, $q=N^{-1} \sum_i S_i^1 S_i^2$.
  \item {\em Trapping times}: the ``trapping time'' associated to a LSC is the time lapse, in MCs, 
  that the system takes to go from the LSC to the maximum connecting it to the next
    LSC, i.e. the time to surmount the barrier.
\end{itemize}

In \cite{Stariolo2019} the aim was to compare results from the single-spin-flip Monte Carlo dynamics
in the $p$-spin Model with the predictions of Bouchaud's Trap Model~\cite{Bouchaud1992}. One of the
important differences
between both models is the fact that traps in the Trap Model are statically and dynamically uncorrelated,
while configuration energies in the $p$-spin Model are correlated random variables.
%, as seen in Sec.~\ref{sec:pspin}. 
%In the limit $p \to \infty$ the energies in the $p$-spin Model become
%uncorrelated, which defines the REM. For generic values of
%$p$, correlations between pairs of configurations depend on the overlap $q$, as given by
%Eq.~(\ref{eq:enercorrels}).
Because of this, in this study of the $p=3$ model, besides considering the actual
sequence of LSC, we also considered sequences restricted to have a maximum overlap  $q_{max}$
between consecutive pairs. Note that in this way, a subset of the actual sequence of LSC is filtered.
Of particular interest is the case $q=0$, with a strict equality,
in which consecutive pairs of LSC are uncorrelated.
This choice was done with the
aim of approaching one of the defining features of the Trap Model, that is, uncorrelated random traps. 
In the other cases, some negative correlations are present, but their weight is negligible.

\section{Results}
\label{sec:results}

\begin{figure}[ht!]
\centering
\hspace*{-0.5cm}\includegraphics[scale=0.75]{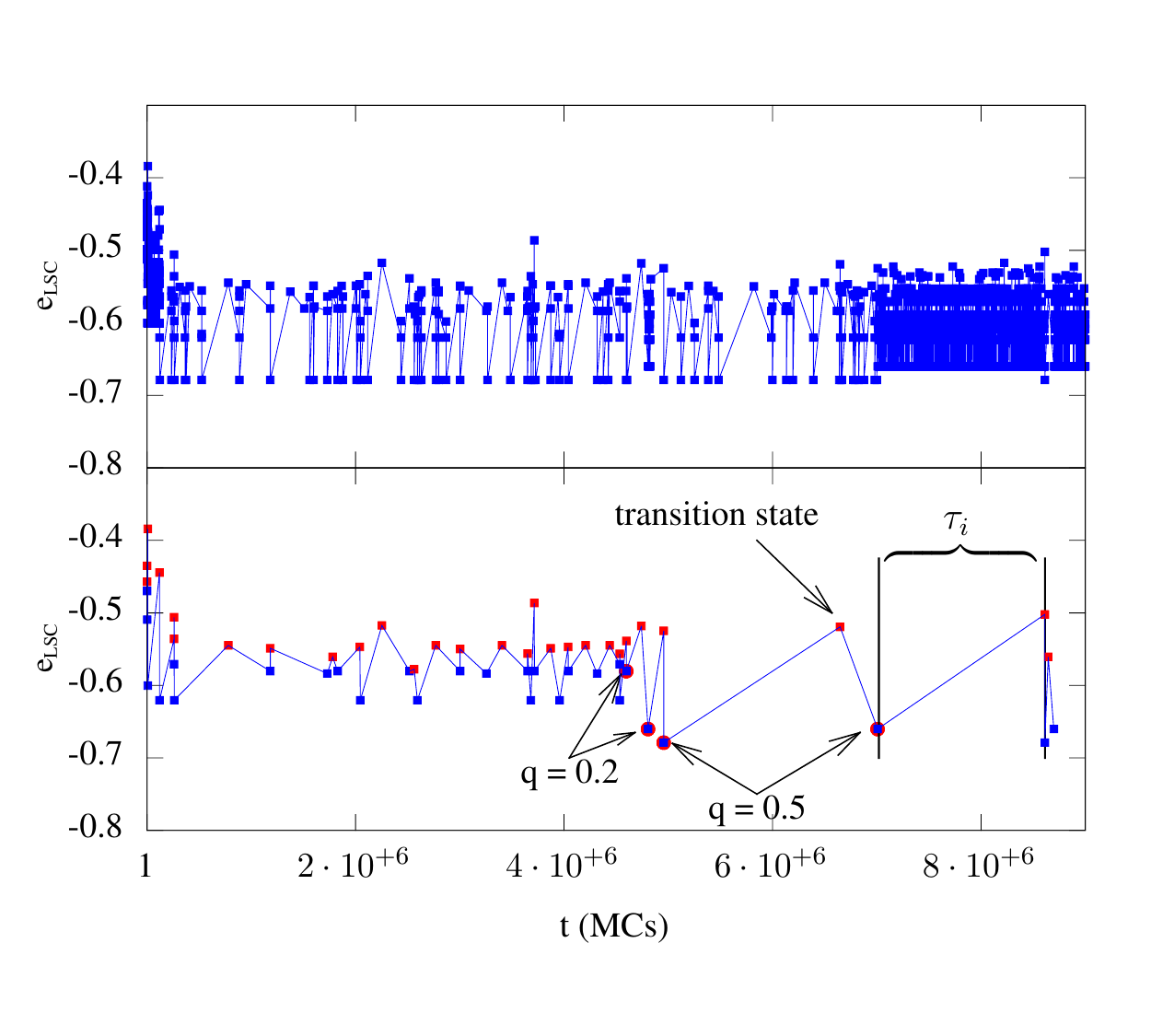}
\vspace{-0.85cm}
\caption{(Color online) Sequence of Locally Stable Configurations (LSC) and the maxima separating them. Top:
  consecutive LSC restricted to have overlap $q < 1$. Bottom: only consecutive LSC
  with $q<0.6$ are shown. Two typical pairs and the corresponding overlaps are highlighted.  For reference, 
  the equilibrium energy density is approximately $-0.69$.}
  \label{fig:singlerun}
\end{figure}

In Fig.~\ref{fig:singlerun} we show the energies of a sequence of LSC and maxima along a typical
quench from a disordered initial state to $T=0.2$ in a system with $N=20$.
In the top panel a sequence of LSC with $q<q_{max}=1$ is shown. 
We note that there are some energy levels which are repeatedly visited by the system.
In the bottom panel only pairs
of consecutive LSC with $q < q_{max}=0.6$ are shown for the same sequence of random numbers.
Two pairs of LSC with $q=0.2$ and $q=0.5$ are highlighted.
The maxima between them are the transition states, one of which is indicated with a legend
(as already mentioned, not to be confused with the transition states usually defined in the context of continuous energy
landscapes, in which they are saddles of index one connecting two local minima).
The trapping time $\tau_i$ of a pair of LSC is also shown.
These plots represent one dimensional snapshots of the $p$-spin energy landscape during the quench.
We can note several characteristics that are present in almost every
instance of the quench dynamics:
\begin{itemize}
\item The sequence of energies of the LSC along a trajectory is not monotonically decreasing.
  In other words, the gaps can be of either sign.
  \item The sequence of maxima between pairs of LSC is also non monotonic.
\item Lower energies do not always imply longer trapping times. 
\item The time of descent from a maximum to the next LSC is not necessarily shorter than the
  trapping time, i.e., the time to go up from the previous LSC to the maximum.
  This is in sharp contrast with the usual relaxation over a simple barrier in a double well potential,
  in which the time of decay from the transition state is negligible in comparison to the time
 needed to reach the transition point. It is a manifestation of the roughness of the
 large dimensional  energy landscape of the $p$-spin Model.
\end{itemize}

\begin{figure}[ht!]
\centering
\includegraphics[scale=0.7]{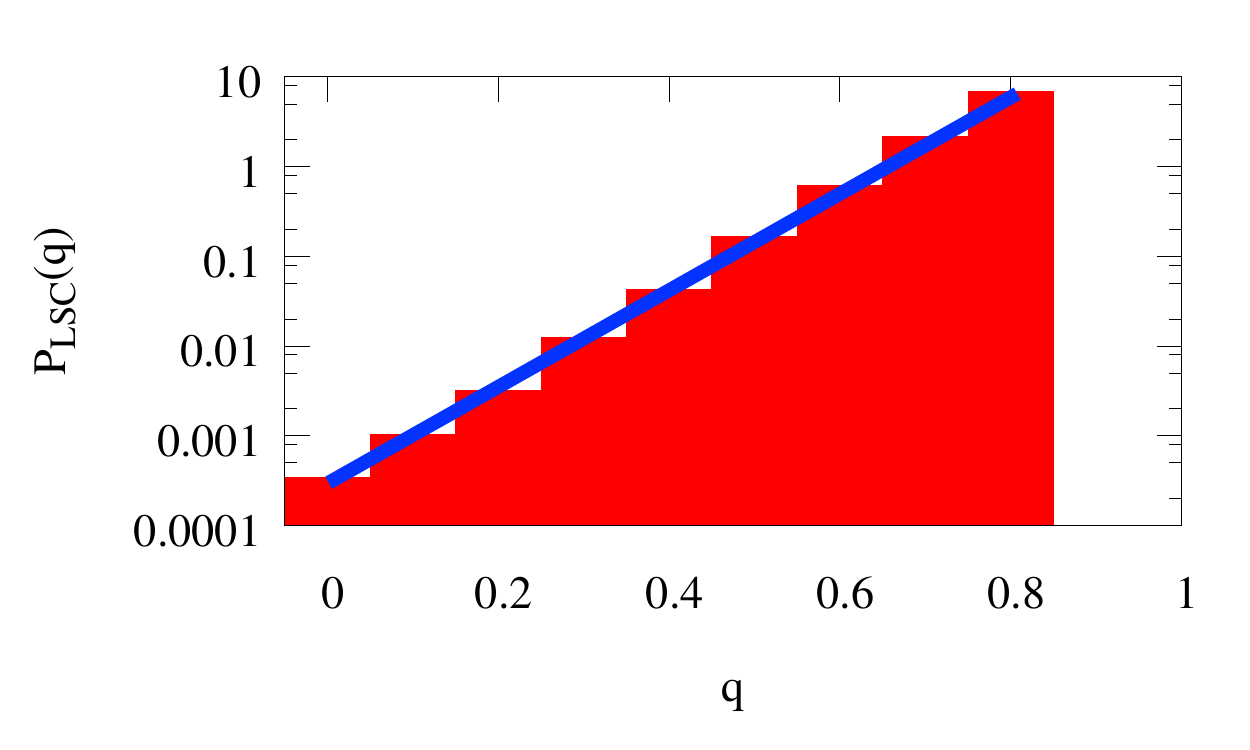}
\caption{(Color online) Histogram of overlaps between consecutive Locally Stable Configurations along
  a Monte Carlo trajectory. Consecutive states with $q=1$ were excluded and $q=0.9$ is not allowed
  by the condition of single-spin-flip stability in a system with $N=20$. }
\label{fig:overlaps}
\end{figure}

Figure \ref{fig:overlaps} displays the normalized distribution of overlaps between consecutive pairs of
LSC. Pairs with overlap $q=1$ were discarded. This was motivated by the fact that there are frequent
situations in which the system visits repeatedly a single LSC, with short excursions to higher energy
nearby states.
By excluding consecutive pairs with $q=1$ we automatically consider the whole sequence in these cases
to be part of
the same trap. It can also be seen that pairs with $q=0.9$ are absent too. This is not imposed, but is a
consequence of the definition of single-spin-flip-stable configurations. In a system with $N=20$,
two configurations differing by a single spin flip will have $q=0.9$. Then, if a configuration is stable
against any single spin flip it cannot move to another stable configuration with one spin being different.
The figure shows an exponential growth of the probability with $q$.
$90\%$ of the weight corresponds to highly correlated pairs with $q \in [0.7,0.8]$.

\begin{figure}[ht!]
  %\centering
  \hspace*{-3.5cm}\includegraphics[scale=1.2]{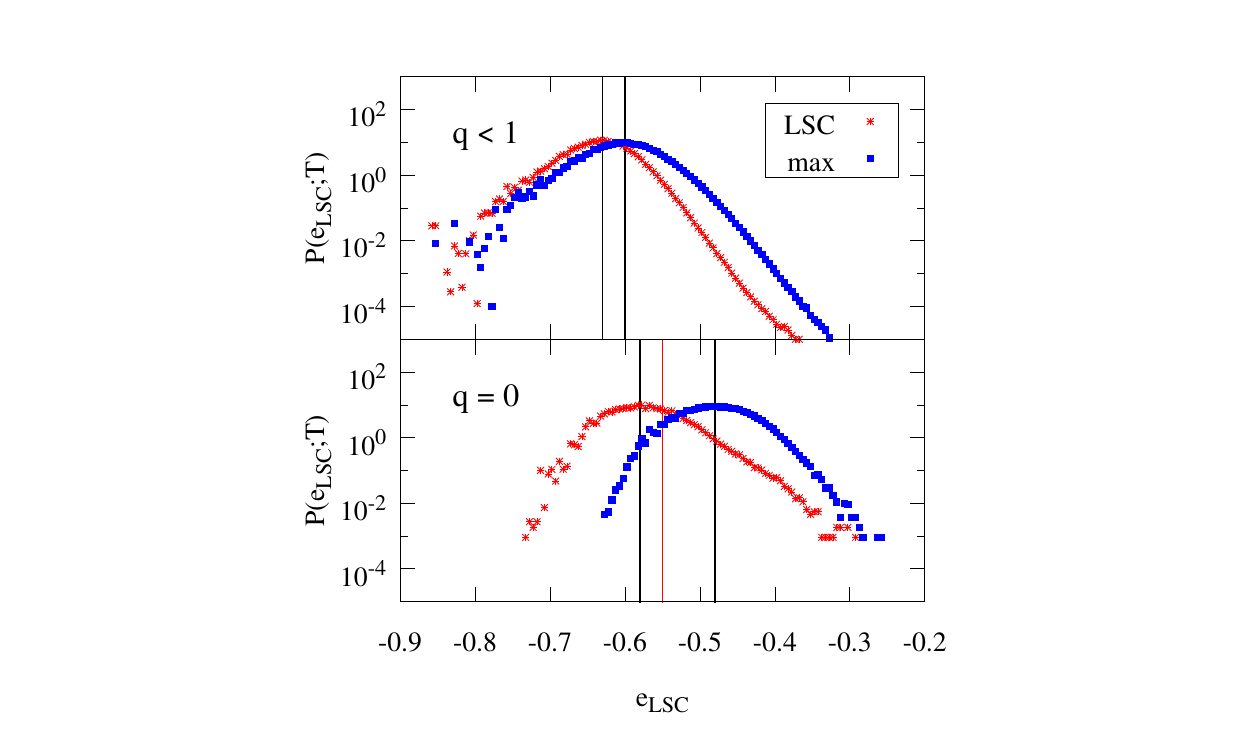}
\caption{(Color online) Probability densities of energies of Locally Stable Configurations (red) and 
 maxima between
  pairs of consecutive LSC (blue) along a Monte Carlo trajectory. The vertical 
  black lines indicate the most probable values and in the bottom panel also the threshold level
  in red (see text).
  Top: configurations with $q < 1$. Bottom: configurations with $q=0$.  }
\label{fig:energies}
\end{figure}

\begin{figure}[ht!]
\centering
\includegraphics[height=.25\textheight,width=.5\textwidth]{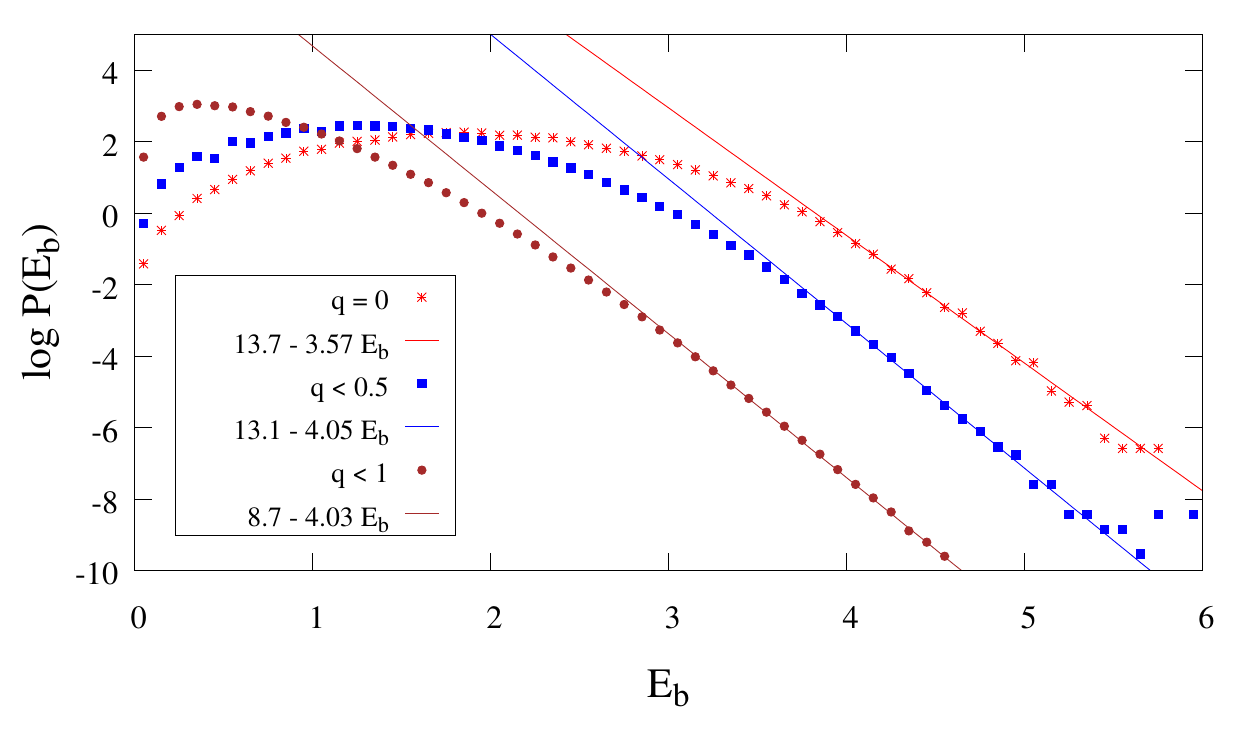}
\vspace{-0.8cm}
\caption{(Color online) Distributions of barrier heights. The tails are ordered from left to right: $q < 1$, $q<0.5$ and $q=0$.
The straight lines are exponential fits to the tails with parameters given in the key.}
\label{fig:barriers}
\end{figure}

Figure \ref{fig:energies} shows distributions of energy densities of LSC together with the
distributions of maxima connecting pairs of consecutive LSC. Two characteristic cases
are shown. In the top panel the distributions correspond to pairs satisfying $q<1$, i.e. nearly
all pairs in a trajectory are included, excepting only those cases in which two consecutive
LSC were the same. 
The bottom panel shows the opposite case, in which $q=0$, i.e.
sequences of uncorrelated LSC. With such a restriction one picks a small subset of the actual
sequence of LSC. Sequences of uncorrelated traps are an essential ingredient of Bouchaud's
Trap Model~\cite{Bouchaud1992,BouchaudDean1995}.
We note that the typical
barriers, i.e. the difference between the most probable minima and the most probable maxima
 are larger in the $q=0$ case. In this case
the system has to climb to higher energy levels in the landscape in order to connect with an
uncorrelated state. Meanwhile, the small typical barriers in the case $q<1$ reflect a flatter
landscape, with mild undulations connecting typical local minima.
A third vertical line at  $e_{th} \sim -0.55$ is shown in the $q=0$ panel.
  This is a finite size ``threshold level'',
computed for the system with $N=20$ by a method proposed in \cite{Hartarsky2019}. Note that it
is in between the maximum of the LSC energies and the maximum of the distribution of maxima.
Looking at the distribution of barriers in Fig.~\ref{fig:barriers}, for $q=0$ we can see that
the relevant energy densities lay in the range
$[0.2;0.25]$ ($E_b$ between 4 and 5 in Fig.~\ref{fig:barriers}).
Then, for getting a barrier height in this range, the
energies of the LSC and corresponding maxima must be at the left and right of $e_{th}$
respectively. In turn, this means that in order to reach the next LSC along the dynamical path,
the system must typically overcome the threshold level. 

\begin{figure}[ht!]
\centering
\includegraphics[height=.2\textheight,width=.5\textwidth]{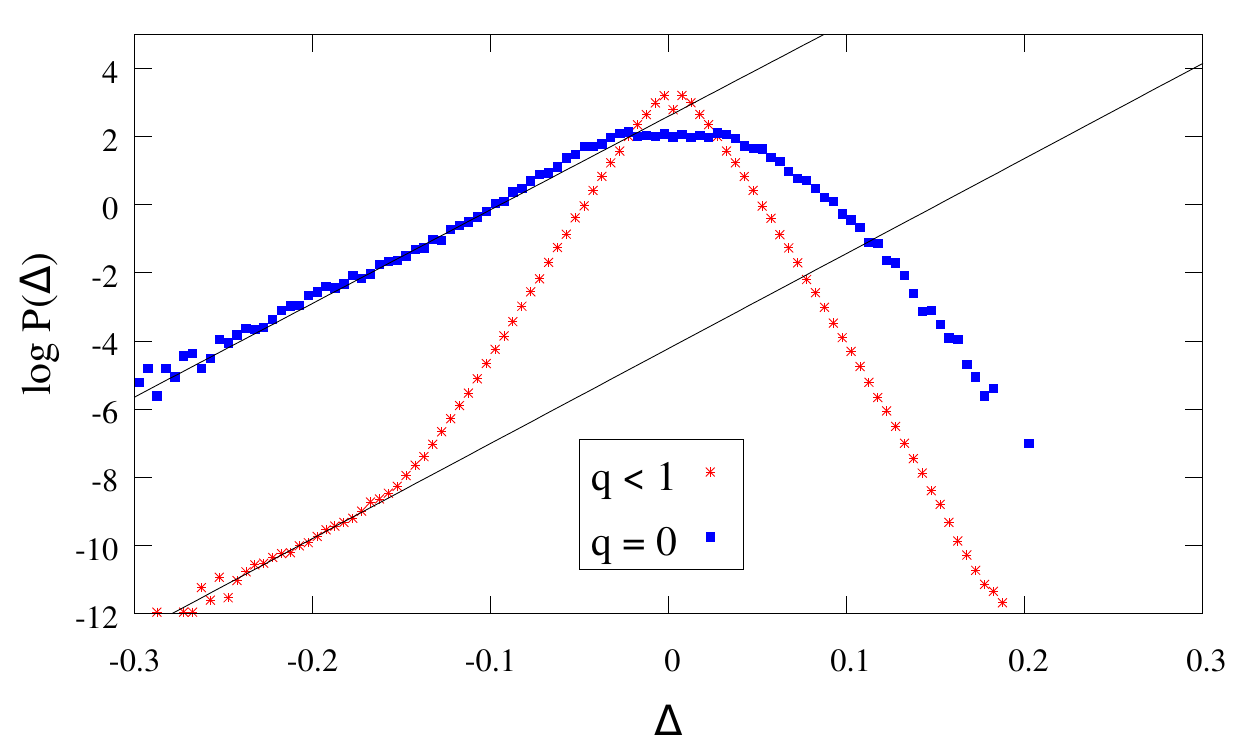}
\caption{(Color online) Distributions of energy gaps between consecutive Locally Stable Configurations. 
Exponential  fits shown in solid lines have the same slope.}
\label{fig:gaps}
\end{figure}

In Fig.~\ref{fig:gaps} the distributions of gaps are shown. A first observation is that both
positive and negative gaps are present in the two cases studied. This means that the
observed processes do not necessarily imply a monotonic descent in energy along the LSC sequences.
One can also note qualitative differences in the cases $q<1$ and $q=0$. In the $q<1$
case, and for relatively small gaps,  the distribution is pretty symmetric, with similar exponential 
regimes both for the positive and the negative sides. This means that starting from a particular LSC it is
equally probable to end in another one at a  higher or lower energy level.
A different regime with a slower exponential rate can be seen at large negative gaps.
Instead, the
distribution for $q=0$ is asymmetric, with larger weight on the negative gap side, meaning higher
probability to go down in energy. In this case an exponential tail is only seen on the negative side
of the distribution.
Fits to the far negative sectors give the same rate in both cases, signalling that
the slower exponential regime in the $q<1$ case corresponds to $q=0$ pairs, and that this set of
uncorrelated pairs behaves differently from all other correlated cases. 

In Fig.~\ref{fig:barriers} we show the corresponding distributions of barrier heights ($E_b=Ne_b$)
for three cases $q<1$, $q<0.5$ and $q=0$.
In all cases an exponential regime in the large barriers sector is observed. Fits to this regime allow
one to obtain the rate of the exponential decay, which will be considered later when confronting these
results with predictions from different models.
The fitting ranges were chosen considering the widest interval on which
  the data points are almost perfectly aligned at eye view. This criterium guarantees the minimum asymptotic
  standard error in the regression fits, shown in Table \ref{tab:fits}
  ~\footnote{The plots showing probability distributions do not show statistical error bars for single
    points. Then, a common goodness-of-fit estimator like the $\chi^2$ fails to give a meaningful number
    in this case, because it cannot assign weights to data points. We
    prefer to show the asymptotic standard error of the fit parameters,
    which corresponds to the standard deviation of the linear least-squares problem.}.
For the two correlated cases, $q<1$ and $q<0.5$, the rate of the exponential decay is
nearly the same, while a definitely smaller rate was obtained in the uncorrelated case, $q=0$.
In the next section we will analyze these results in connection with Trap Models.

\begin{figure}[ht!]
\centering
\includegraphics[height=.25\textheight,width=.5\textwidth]{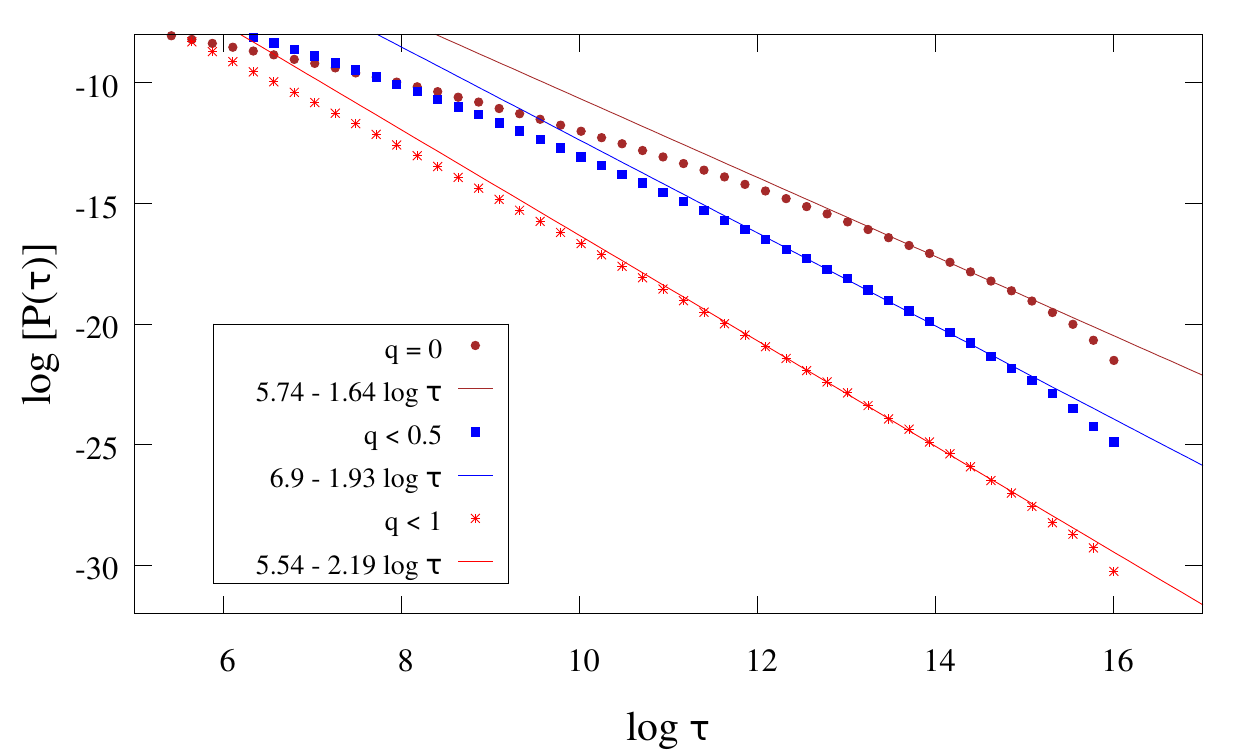}
\caption{(Color online) Distributions of times to reach the maximum between consecutive locally
  stable configurations. From bottom to top: $q < 1$, $q<0.5$ and $q=0$. The parameters of the 
  exponential fits shown with thin straight lines are given in the key.}
\label{fig:times}
\end{figure}

Figure \ref{fig:times} shows the distributions of trapping times. We note that the slopes of the
curves are slowly changing with the time scale $\tau$.
We will be interested  in the largest trapping times.
  These are limited by the time span of the simulation.
Then, it is natural to expect finite time effects for the longest times due to insufficient statistics.
We have run simulations with different total Monte Carlo steps (not shown) and verified that, 
because of the algebraic growth in the measuring times, the last 4 or 5 points are affected by finite
time effects. In Fig.~\ref{fig:times}, the last 4 points correspond to the second half of the total time
span of the simulation and, consequently, for each disorder sample at most one such long trap time can
be observed. Because of this, we decided to discard the last four points from the fitting ranges.
The results for the fits together with the corresponding asymptotic standard errors are shown in Table \ref{tab:fits}.

In the next Section we will remind the definition of a few well-known models of aging dynamics with
activation mechanisms, which we generically  call Trap Models. They will be the reference frame
for analysing the results on the $p$-spin Model that we have presented in this Section.

\section{Trap Models}
\label{sec:comparison}

In the original Bouchaud's Trap Model (BTM) the system is defined by an infinite set of configurations,
or ``traps'',
with energies $E_i$ that are i.i.d. random variables chosen from an
exponential pdf given by~\cite{Bouchaud1992,BouchaudDean1995}:
  \beq
P(E_i) = \lambda \ \exp{[\lambda\,(E_i-E_0)]} \; , \hspace{1cm} E_i \leq E_0
\; ,
\label{eq:exp-dist}
\eeq
where $1/\lambda$ is the mean and $E_0$ is a reference energy level, usually chosen to be
zero. The system stays confined in a trap of energy $E_i$ during  a ``trapping time'' which is also an
exponentially distributed random variable with average $\tau_i = \tau_0 \ \exp{[\beta \,(E_{0}-E_i)]}$,
where $\beta \equiv 1/T$ is the inverse temperature. After escaping from a trap the system can jump
to any other one with equal probability. This implies a renewal character of the dynamical process:
after escaping from a trap the dynamics restarts anew and memory of the past history is lost. This
property allows the stochastic dynamics to be solved exactly. One of the main outcomes
of the solution is the distribution of mean trapping times:
\beq \label{eq:trap-times}
\phi(\tau;x) = x\  \frac{\tau_0}{\tau^{1+x}}
\; ,
\eeq
where $x=\lambda /\beta=T/T_c$. For $T<T_c$ the average trapping time diverges and the system
ages forever. Another distinctive characteristic of the BTM dynamics is the behaviour of the two-times
correlation function $C(t_w,t_w+t)$, which is given for long $t_w$ and long $t$ by the so-called
``Arcsin law''~\cite{BouchaudDean1995}.
The ubiquity of the Arcsin law in different models related with the BTM has been
extensively studied, specially in connection with the REM dynamics
~\cite{Arous2002,BenArous2003,Bovier2003,Gayrard2016,Cerny2017,Cammarota2015,Cammarota2018,Baity-Jesi2018a}.
Note that in the REM, as well as in the more general $p$-spin Models,
the energies are Gaussian i.i.d. random variables.
In particular, it has been shown that the long time auto-correlation function of the REM can be
mapped exactly on the (exponential) BTM Arcsin law, provided the times of observation are scaled with a
factor that grows exponentially with system size. More precisely,
\beq
\lim_{\stackrel{t_w \to \infty}{t/t_w=\omega}}\lim_{N \to \infty} C(\theta(N)t_w,\theta(N)(t_w+t)) = H_x(\omega)
\eeq
where
\beq
H_x(\omega)=\frac{\sin{(\pi x)}}{\pi}\int_{\omega}^{\infty} \frac{du}{(1+u)\,u^x}
\; ,
\eeq
and $\theta(N)=\exp{(\gamma N)}$ is the time rescaling factor. In a model with $M=2^N$ configurations,
the dynamics can be seen as an exploration of sets of $2^{\rho N}$ configurations, with $0<\rho <1$.
In the context of the Gaussian Trap Model, a Trap Model with a Gaussian density of states, 
it can be shown that $x=\sqrt{\rho}\; T/T_c$, and
$\gamma=x/T^2$~\cite{Arous2002,Diezeman11,Cammarota2018}. As $\rho$ grows with time, the previous results imply
that the $x$ exponent in the distribution of trapping times {\em depends on the time scale of
  observation}.

Coming back to the results presented in Sec.~\ref{sec:results}, and with the aim of making a quantitative comparison with
the BTM predictions, we considered that the barrier heights in the $p$-spin Model, and not the energies of the
LSC, are the analogs of the trap energies in the BTM. The reason for this is the following: in the BTM, because there is a
unique energy level $E_0=0$ to which the system has to arrive in order to jump to another trap, the
energy depths are in fact energy barriers. At variance with this characteristic of the BTM, in the
$p$-spin Model the transition states appear at different energy levels. As seen before, the pdfs of
Fig.~\ref{fig:barriers} show an exponential regime in the large barriers sector. 
The results of the fits for the exponential rate $\lambda$ are shown in Table \ref{tab:fits}
for $T=0.2$. We note that the result is the same for the two correlated cases while it is different
in the uncorrelated one.

Now, one can use these rates to obtain predictions for the exponent $1+x$ of the trapping times pdfs
for different models for which exact results are known. The results for the BTM prediction
$1+x = 1+\lambda/\beta$ are shown in the second column of Table \ref{tab:fits}.
These values have to be compared with the numerical results for the trapping
times distributions. Looking at the curves in Fig. \ref{fig:times} one notes a drift in the slope
of the distributions, meaning different time scales are probed by the dynamics,
in agreement with the observation in the context of Trap Models with Gaussian distributed energies.
The long time regime, corresponding to the last 2 or 3 decades in the figure,
may be expected to correspond to the exponential regime of large barriers in Fig.~\ref{fig:barriers}.
The results of the fits for the exponent $1+x$ are shown in the last column of Table \ref{tab:fits}.
The values from the fits do not match the BTM estimates based solely on the distribution of
barrier heights and the temperature. Having said this, we also observe that the discrepancy 
diminishes as the correlation between consecutive LSC is reduced.

\begin{widetext}
\begin{center}
  \begin{table}[!ht]
\begin{tabular}{||c|c|c|c|c|c||}
  \multicolumn{6}{c}{}\\ \hline
  \em{}   &  Fits  & \em{Trap Model}   & \em{$a$-generalized TM} & \em{Step Model} & Fits \\ 
  \em{}   & $\lambda$ & $ \; 1+x=1+\dfrac{\lambda}{\beta} \; $ &  $ \; 1+x=1+\dfrac{\lambda}{(1-a)\beta} \; $ & 
  $ \; 1+x=3-\dfrac{\beta}{\lambda} \; $ & $\; 1+x \; $ \\ \hline
  $\; q< 1.0\; $  &  \; 4.03$\pm$ 0.04  \; & \; 1.81$\pm$ 0.01  \; & \; 2.60$\pm$ 0.01 \; & \; 1.76$\pm$ 0.01  \; & \; 2.19$\pm$ 0.02 \;  \\ \hline
  $\; q< 0.5\; $  &  \; 4.05$\pm$ 0.08  \; & \; 1.81$\pm$ 0.02 \; & \; 2.60$\pm$ 0.02 \; & \; 1.77$\pm$ 0.03  \; & \; 1.93$\pm$ 0.02  \; \\ \hline
  $\; q=0 \; $     &  \; 3.57$\pm$ 0.08  \; & \; 1.71$\pm$ 0.01 \;  & \; 2.43$\pm$ 0.01 \; & \; 1.60$\pm$ 0.03 \; & \; 1.64$\pm$ 0.03  \; \\ \hline
\end{tabular}
\caption{Rates $\lambda$ of the exponential regime of the barrier height distributions
  and power law exponents $x$ of the trapping times pdfs
  from fits to the results of Figs. \ref{fig:barriers} and
  \ref{fig:times}, $\beta=1/T=5$, together with predictions for two variants of the Trap  and Step Models.
For the $a$-generalized TM the numerical results were obtained for $a=1/2$.}
\label{tab:fits}
  \end{table}
  \end{center}
\end{widetext}

As discussed above, one of the differences between the original BTM and more general models with rough
energy landscapes, is that in the BTM the system always has to jump to a fixed energy level,
a threshold $E_{th}=E_0$, in order to escape from any trap. This is usually referred to as a ``golf course''
Trap Model landscape.
Generalizations in which the trapping times depend on both
the initial $E_i$ and final $E_j$ energies have been considered in 
Refs.~\cite{Rinn2000,Rinn2001,Monthus2003,Cammarota2018}. For discrete dynamics on the
$N$-dimensional hypercube, as relevant to the $p$-spin Model single-spin-flip dynamics, the
``$a$-generalized'' transition rates
from an initial state with energy $E_i$ to a final configuration with energy $E_j$
read~\cite{Rinn2000,Rinn2001,Cammarota2018}:
\beq \label{eq:ageneral}
r_{i,j} \propto \exp{\left[\beta(1-a)E_i-\beta a E_j\right]}
\; ,
\eeq
where $a \in [0,1)$. In the case $a=0$ the dynamics are equivalent to those of the original BTM.
In the case $a=1/2$
both levels have the same weight in the transition rate, similar to a Metropolis dynamics with a
temperature that is double the usual one. Within these $a$-generalized dynamics the average
trapping times for a threshold level $E_0=0$ are given by
$\tau_i = \tau_0\,\exp{\left(-\beta(1-a)E_i\right)}$.
Then, if the energies are exponentially distributed with rate $\lambda$, the
exponent of the trapping times pdf should be generalized to $1+x = 1+\lambda/[(1-a)\beta]$.
Applying this reasoning to our results, with $a=1/2$, gives $1+x=2.6$ for the case $q<1$
  and $q<0.5$, and $1+x=2.43$ in the case $q=0$ (see Table \ref{tab:fits}).
These results are even further away from the naive Trap Model predictions. In particular, in the
three cases, $x > 1$ which is not compatible with a system in the full aging regime.
Nevertheless, as noted and discussed in \cite{Cammarota2018}, the $a$-generalized dynamics differ
from the single-spin-flip Metropolis dynamics in a detail that implies a very different nature of the
exploration of the energy landscape. Consider the case of interest here with $a=1/2$. In this case, 
Eq.~(\ref{eq:ageneral}) becomes
$r_{i,j} \propto \exp{\left(-\beta \Delta E /2\right)}$, with $\Delta E= E_j-E_i$. 
This form implies that even the rates of descent from
higher to lower energy levels are dominated by the transition from the highest energy level $E_{max}$
to the minimum one $E_{min}$. All other downward transitions are exponentially smaller.
On the other hand, with the Metropolis rule, see Eq.~(\ref{eq:Metropolis}), or
with the Glauber transition rates,
\begin{equation}
r_{i,j} \propto \frac{e^{-\beta \Delta E}}{1+ e^{-\beta \Delta E}}
\; , 
\end{equation}
any downward transition is accepted with probability one.
At very low temperatures, this provides a microscopic mechanism for lowering the energy {\em without
  the need for activation}. This was first noted in the behaviour of the Step Model
(SM) with Glauber dynamics~\cite{Barrat1995}, in which
the set of $M$ i.i.d. random energy levels $E_i$ are chosen with
a  probability $P(E_i)$ also given by Eq.~(\ref{eq:exp-dist}).
  Similarly to what happens with the Metropolis
dynamics of the $p$-spin Model, the dynamical rule in the SM does not
imply an activation mechanism as a necessary condition for
relaxation, at variance with the BTM. Instead, at very low temperatures
relaxation is governed by an entropic mechanism, i.e. the search for favorable paths in configuration
space to go down in energy. Further work on the SM noted that at intermediate temperatures activation
over barriers is also present, when the probability to go up in energy becomes larger,
and there appears a competition between entropic relaxation and activation,
leading to an effective trap-like
phenomenology, but with a different set of exponents for the distribution of trapping times~\cite{Bertin2003}. 
Nevertheless, in order to observe effective trap-like exponents in simulations
of the SM with single-spin-flip dynamics,
it is necessary to look at coarse-grained time scales~\cite{Cammarota2015}. In fact, while at low relative temperatures
$\lambda/\beta <0.5$ the probability to go down in energy is higher than the probability to go up,
leading to an essentially entropic relaxation, in the regime $0.5<\lambda/\beta<1$ the opposite relation
holds. This induces explorations of high energy levels in the landscape
before relaxation to deep states happens. Furthermore, an energy threshold level $E_{th}$ can be
defined by the condition of equality between the probability to go up and go down, which was used
to set a coarse-grain time scale in the SM~\cite{Cammarota2015}. Then, traps were redefined as portions
of the energy landscape (basins) visited by the Metropolis dynamics while configuration
energies obey $E < E_{th}$. When $E > E_{th}$ the system is considered to be in a transition state, until
it goes down below the threshold again and explores a new basin.
Redefining traps as basins and considering the value of the exponent
$x=2-\beta/\lambda$ for the SM dynamics in the intermediate regime $0.5<\lambda/\beta<1$, it was possible
to verify the validity of the BTM paradigm for aging dynamics also in this case where competition
between activation and entropic relaxation is at work~\cite{Bertin2003,Cammarota2015}. 
It is then tempting to compare our results on the Metropolis dynamics of the
$p$-spin Model, with the predictions for the SM, although the protocol used in the present work does not
correspond to the definition of basin in the SM.
Note that basins
in the SM are equivalent to states or energy levels in the BTM. The relevant piece of information
to obtain the trapping times (and their distribution), while it is associated to a basin or to a single state,
is a barrier height defined by a threshold level. In our approach
for the $p$-spin Model we do not make reference to a fixed threshold level.
Instead, we considered as relevant the fact that, in order to go from a LSC to a nearby one,
the system must climb a barrier which depends not only on the static landscape, as is the case in the
BTM and SM, but on the actual dynamical path. Because of this, we think that barrier heights in the
$p$-spin Model play the role of energies of the BTM or the SM.

For $q<1$ and $q<0.5$ we obtained the ratio $\lambda/\beta = 0.8$, while for $q=0$ it was $\lambda/\beta=0.71$,
  both corresponding to the intermediate regime of the SM, $0.5<\lambda/\beta<1$.
  When comparing the exponent of the
  trapping times distribution using the results of the fits for the barrier heights we deduce
  $1+x=3-\beta/\lambda=1.75$ for $q<1$ and $q<0.5$, which does not agree
  with the values $1+x\approx 2.19$ nor with $1+x \approx 1.93$ inferred from the direct fit to
  the distribution of trapping times for $q<1$ and $q<0.5$ respectively (see Table \ref{tab:fits}).
  For the case $q=0$ we obtain $1+x=3-\beta/\lambda \approx 1.60$ using the value of the rate $\lambda$
  from the fit to the barriers distribution, and $1+x\approx 1.64$ from a direct fit to the trapping
  times pdf. 
Interestingly, we observe a reasonable agreement, within numerical uncertainties, between the
prediction of the exponent $x$ for a
SM with exponential energies and our $p$-spin results based on sequences
of uncorrelated LSC. Still, for the case of correlated sequences, the results do not show agreement. 

\section{Discussion}
\label{sec:discussion}

We analyzed the dynamics of the Ising $p$-spin Model with single-spin-flip Metropolis updates
from the perspective
of the Trap Model paradigm. The quantitative predictions of the BTM rely on
the static independence of the trap energies and also on the dynamic independence between consecutive
visited traps. These two properties lead to the well known power law distribution of mean trapping times
and to the Arcsin law for the two-times correlation function in the aging regime, when $x=\lambda/\beta <1$.
None of the previous two defining ingredients of the BTM are present in the Ising $p$-spin with
Metropolis dynamics. Another difficulty for testing the predictions of Trap Models against systems with
single-spin-flip dynamics is the very definition of trap. While traps in the original BTM are single
energy levels, in the $p$-spin and related models this has proven not to be true. 

With these difficulties
in mind, we decided to follow trajectories of the $p$-spin Model looking at sequences of configurations
which are stable against single spin flips, or locally stable configurations (LSCs). Because they show
some degree of stability, they seem a priori good candidates to act as trapping configurations of the
dynamics. Nevertheless, they are not statically nor dynamically independent in general. Accordingly, we decided
to study sequences of LSCs depending on the degree of correlation between pairs of successive
configurations. We have seen, as shown in Fig.~\ref{fig:overlaps}, that most of these configurations are
strongly correlated. Then, we focused on three representative situations, when pairs of successive LSCs
are restricted to have overlaps $q<1$, $q<0.5$ and $q=0$ respectively, and we compared results for the
distribution of energy barriers (that we took as similar objects to the trap energies in the BTM) 
and trapping times, with the predictions from a set of models 
that, in recent studies, have been shown to conform to the Trap Model paradigm.

A first outcome of our study is the fact that, at least at low temperatures, the pdfs of barrier heights
from LSCs show an exponential regime for large barriers, seen in Fig.~\ref{fig:barriers}. A look at the
distribution of energies and maxima of Fig.~\ref{fig:energies} allows one to point out that,
although the energies of
the LSCs visited by the dynamics with $q<1$ and $q=0$ restrictions are similar, in the latter case the system has to climb
higher in the landscape in order to decorrelate. 
The presence of an
exponential regime of the barrier heights at large values allows one to try a direct comparison with
different Trap Model expectations.

Our second piece of information comes from the pdfs of trapping times. They show a slow decrease of
slope for increasing trapping time, which is compatible with the expected behaviour of Gaussian Trap Models.
Then, one should compare the results from the
exponential regime of barrier heights with the large times sector of the trapping times distributions.
As shown in Table \ref{tab:fits}, the numerical results for $1+x$ are around $2.0$ in the correlated cases,
too large according to the expectations of the naive BTM,
and also different from the result coming from the distribution
of barrier heights, $\approx 1.8$, in both correlated cases. In the $q=0$ case,
 the numerical values $1+x\approx 1.64$ from the trapping times pdf, and $1+x\approx 1.71$ from
the barriers pdf are nearer to each other, although it does not seem to be possible to get much more
precise numerical results.
Comparison with the $a$-generalized Trap Model predictions are still worse. Choosing $a=1/2$, in order to
give equal weights to the initial and final states in the dynamical rule,
 we obtain $1+x$ exponents
larger than two in all correlated and uncorrelated cases, not compatible with an aging dynamics.

\begin{figure}[h!]
  \centering
  \includegraphics[height=.25\textheight,width=.5\textwidth]{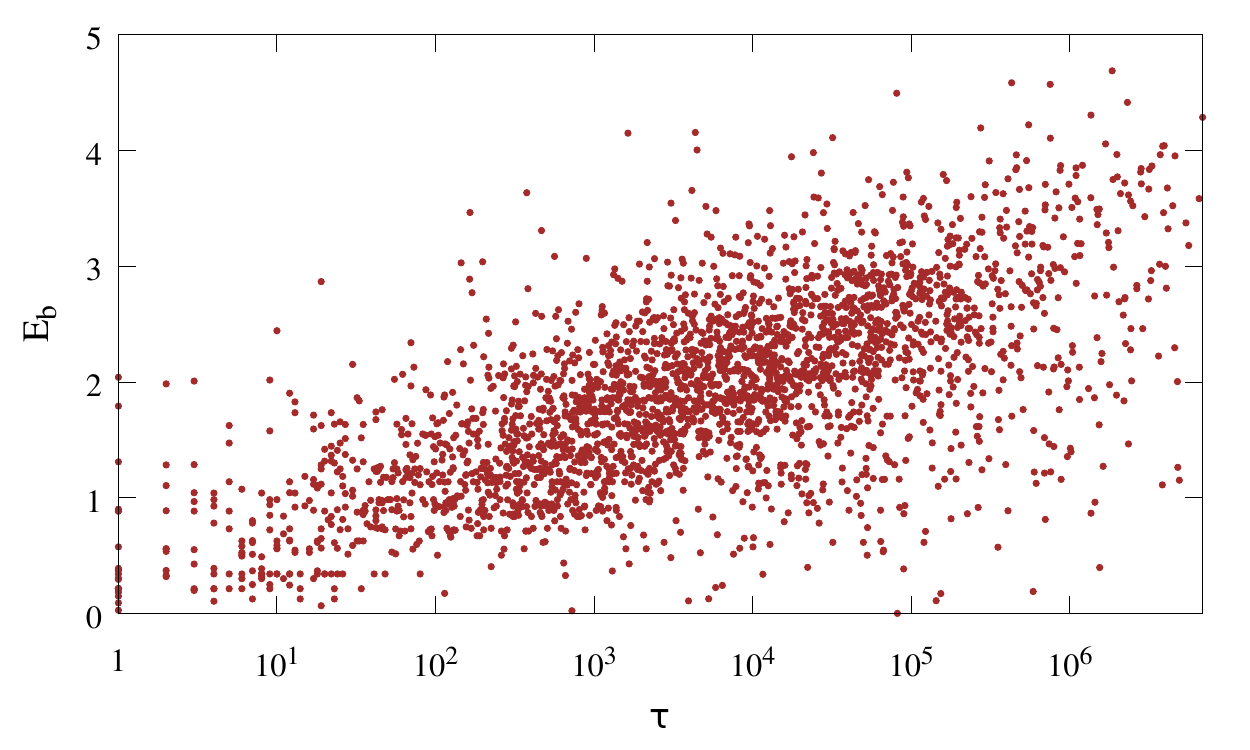}
\caption{(Color online) Scatter plot of barrier heights versus trapping times for $N=20$ and $T=0.2$.}
\label{fig:scatter}
\end{figure}

Better results are obtained when comparing the numerical ones with the predicitions of the Step Model.
In this case, while the two correlated cases do not show quantitative
agreement between trapping times and barrier height predictions, in the uncorrelated case surprisingly
close results are obtained, with $1+x \approx 1.64$ from a fit to the trapping times pdf and
$1+x\approx 1.60$ using the analytic result $1+x = 3-\beta /\lambda$, with the $\lambda$ value obtained
from the barriers pdf. This is interesting, because in principle it is unexpected. Although considering
uncorrelated sequences of configurations brings us a step closer to one of the basic properties
of the Step Model, as also does the Metropolis dynamics, 
our present trap  definition 
is different from the one considered, e.g. in the numerical study in Ref.~\cite{Cammarota2015}. As discussed
above, in that study traps were defined as sets of configurations, or basins, separated from each
other during the dynamics whenever the energy reached a threshold level $E_{th}$, defined by the condition
of equality between the probability to go up or down in a single time step. In the $p$-spin case, we
have not been able to define a similar threshold level for the small size systems considered in this
work (although a similar approach can be pursued following the observations made when discussing relevant
energy scales in Fig.~\ref{fig:energies}).
Instead, we defined traps relying on pairs of configurations with restrictions on the
overlap along the dynamical path.
On the other
side, the closer correspondence between the $p$-spin results and the Step Model,
does not come as a surprise when one considers the physical
mechanisms for relaxation. In Trap Models relaxation is purely activated. There are no downward dynamical
paths without energy cost. Meanwhile, the Glauber or Metropolis rules in the SM and in the $p$-spin Model induce an
entropic mechanism for relaxation, together with activation, depending on the temperature range.
Further evidence for the presence of a kind of entropic relaxation in the $p$-spin Model can be
  inferred from a scatter plot of raw data showing barrier heights and corresponding trapping times, shown
  in Fig.~\ref{fig:scatter}. Ideally, a BTM behaviour should be seen as a straight line, expressing a perfect
  exponential relation between trapping times and energy. Instead, a dispersion of the data is seen with a clear
  excess density below the ideal straight line. Therefore, large trapping times can be observed with no
need to climb high energy barriers, a typical entropic behaviour.

In the large $N$ limit, it is well known that the $p$-spin Model has an exponential number of saddles of
all indexes, even below the dynamical threshold energy level below which minima exponentially
dominate over saddles of higher index. The precise balance between activation and entropic relaxation
below the threshold level is still an open problem in glassy relaxation. Controlled numerical studies
of the $p$-spin Model with small $N$, instances in which barriers do not diverge in height, 
may be a good starting point in this direction. 

%\noindent{\bf Acknowledgements:}
\acknowledgements
DAS thanks the Laboratoire de Physique Th\'eorique et Hautes Energies at Sorbonne Universit\'e for 
hospitality during the initial steps in the development of this work. DAS was financed in part by
the Coordena\c c\~ao de Aperfei\c coamento de Pessoal de
N\'ivel Superior - Brasil (CAPES) - Finance Code 001 and by CNPq, Brazil.

%\bibliographystyle{apsrev4-1}
%\bibliography{../../ISING-PSPIN-edited}

%

\end{document}